\documentclass[twocolumn,astrosymb]{aastex701}
\usepackage{amsmath}
\usepackage{caption}

\newcommand{\aref}[1]{\hyperref[#1]{Appendix \ref{#1}}}

\begin{document}

\title{Element nucleosynthetic origins from abundance spatial distributions beyond the Milky Way}

\author[0000-0001-7373-3115]{Zefeng Li}
\affiliation{Centre for Extragalactic Astronomy, Department of Physics, Durham University, South Road, Durham, DH1 3LE, UK}
\email[show]{zefeng.li@durham.ac.uk}  

\author[0000-0003-3893-854X]{Mark R. Krumholz} 
\affiliation{Research School of Astronomy and Astrophysics, Australian National University, Cotter Road, Weston Creek, 2611, ACT, Australia}
\email{mark.krumholz@anu.edu.au}

\author[0000-0002-5456-523X]{Anna F. McLeod}
\affiliation{Centre for Extragalactic Astronomy, Department of Physics, Durham University, South Road, Durham, DH1 3LE, UK}
\email{anna.mcleod@durham.ac.uk}

\author[0000-0003-1192-5837]{A. Mark Swinbank}
\affiliation{Centre for Extragalactic Astronomy, Department of Physics, Durham University, South Road, Durham, DH1 3LE, UK}
\email{a.m.swinbank@durham.ac.uk}

\author[0000-0003-1657-7878]{Emily Wisnioski}
\affiliation{Research School of Astronomy and Astrophysics, Australian National University, Cotter Road, Weston Creek, 2611, ACT, Australia}
\email{emily.wisnioski@anu.edu.au}

\author[0000-0002-6327-9147]{J. Trevor Mendel}
\affiliation{Research School of Astronomy and Astrophysics, Australian National University, Cotter Road, Weston Creek, 2611, ACT, Australia}
\email{trevor.mendel@anu.edu.au}

\author[0000-0002-2545-5752]{Francesco Belfiore}
\affiliation{INAF - Osservatorio Astrofisco di Arcetri, largo E. Fermi 5, Firenze, 50127, Italy}
\email{francesco.belfiore@inaf.it}

\author[0000-0002-5281-1417]{Giovanni Cresci}
\affiliation{INAF - Osservatorio Astrofisco di Arcetri, largo E. Fermi 5, Firenze, 50127, Italy}
\email{giovanni.cresci@inaf.it}

\author[0000-0001-8349-3055]{Giacomo Venturi}
\affiliation{Scuola Normale Superiore, Piazza dei Cavalieri 7, Pisa, 56126, Italy}
\email{giacomo.venturi1@sns.it}

\author[0000-0003-2280-2904]{Jialai Kang}
\affiliation{Centre for Extragalactic Astronomy, Department of Physics, Durham University, South Road, Durham, DH1 3LE, UK}
\email{jialai.kang@durham.ac.uk}

\begin{abstract}
An element's astrophysical origin should be reflected in the spatial distribution of its abundance, yielding measurably different spatial distributions for elements with different nucleosynthetic sites. However, most extragalactic multi-element analyses of gas-phase abundances to date have been limited to small numbers of sightlines, making statistical characterization of differences in spatial distributions of elements impossible. Here we use integrated field spectroscopic data covering the full face of the nearby dwarf galaxy NGC 5253 sampled at 3.5-pc resolution to produce maps of the abundances of oxygen, nitrogen, and sulfur using independent direct methods. We find strong evidence for differences in the elements' spatial statistics that mirror their predicted nucleosynthetic origins: the spatial distributions of oxygen and sulfur, both predominantly produced in core-collapse supernovae, indicate that initial injection occurs on larger scales than for nitrogen, which is predominantly produced by asymptotic giant branch stars. All elements are well-correlated but oxygen and sulfur are much better correlated with each other than with nitrogen, consistent with recent results for stellar abundances in the Milky Way. These findings both open a new avenue to test nucleosynthetic models, and make predictions for the structure of stellar chemical abundance distributions.
\end{abstract}



\section{Introduction}

Heavy elements (metals) are synthesized inside stars, ejected from their birth places into the interstellar medium (ISM), then transported and diffused by the motions of ISM gas before being incorporated the next generations of stars. Nucleosynthetic models predict that four physical processes dominate the production of most metals: core-collapse (CC) supernovae (SNe) marking the ends of the lives of stars more massive than $\sim 8$ M$_{\odot}$, asymptotic giant branch (AGB) stars that are the penultimate evolutionary stage for stars less massive than $\sim 8$ M$_{\odot}$, Type Ia supernovae resulting from thermonuclear detonation of degenerate stars, and neutron star merging \citep{Johnson19, Kobayashi20a}. To date, this theory has not been confirmed by observations beyond the Milky Way, but here we investigate a new approach to doing so by investigating whether the origins of elements can be inferred from their abundance distributions, and, conversely, whether the chemical abundance distributions of elements from different origins show quantitative differences. Addressing these questions is crucial for understanding how elements are synthesized and distributed at galaxy scales.

The three metals most readily measurable in interstellar gas (as opposed to in stars), oxygen, nitrogen, and sulfur, are predicted by theory to have different origins: oxygen and sulfur are primarily produced through CC SNe, while nitrogen is dominated by the AGB channel, with CC SNe as a sub-dominant additional contributor. There have been numerous efforts to measure nitrogen-to-oxygen (N/O) ratios in the ISM, but most have involved pencil-beam surveys targeting a small number of H~\textsc{ii} regions or only small portions of a galaxy \citep{Berg15a, Arellano-Cordova16a, Kumari18a, Dominguez-Guzman22a, Arellano-Cordova24a, Gao24}. Such studies can address the typical level of variation in the N/O ratio in such regions, but cannot compare the spatial distributions of these two elements in detail \citep{Kobulnicky97}. By contrast, analyses of the spatial statistics of the oxygen distribution from abundance maps covering the faces of nearby galaxies provide a powerful tool for studying chemical enrichment and transport \citep{Kreckel20, Li21, Metha21a, Metha22a, Williams22, Li23, Li24c}. These methods have recently been deployed for nitrogen as well \citep{Bresolin25}, but only in data where the nitrogen abundances are not independently measured but are instead inferred from the oxygen abundances. Moreover, they have yet to be applied to any other elements, including sulfur. The primary difficulty in doing so is that wide-coverage abundance maps are easiest to obtain using so-called ``strong-line'' calibrations that implicitly assume correlations between N and O and thus preclude measuring the actual correlation of the spatial patterns of the two elements \citep{Kewley19a}. Independent measurements require the use of direct (electron-temperature) methods, which in turn rely on mapping faint temperature-sensitive auroral lines at substantial observational expense.

Here we overcome this obstacle to obtain direct and highly spatially-resolved measurements of the O, N, and S abundance patterns in NGC 5253, a nearby star-forming blue compact dwarf galaxy. This target is advantageous because it is nearby, metal-poor, and has intermediate inclination -- all factors that increase the weak signal of the auroral lines (see \autoref{sec:data} for details). NGC 5253 was observed by the Multi-Unit Spectroscopic Explorer \citep[MUSE;][]{Bacon10} integrated field unit (IFU) spectrograph at the Very Large Telescope (VLT). The data (Program ID: 1104.A-0026E; PI: Wisotzki) are included in the DWALIN survey \citep{Marasco23a}, which serves as our main data set as MUSE covers all the essential emission lines used in O, N, and S abundance measurements. We report the spatial auto- and cross-correlations of O, N, and S abundance maps, offering a detailed understanding of element nucleosynthesis origins. 

\section{Observations, sample selection, and data reduction}
\label{sec:data}

We select our target galaxy from the DWALIN dataset \citep{Marasco23a} that comprises 40 nearby star-forming dwarf galaxies with low stellar masses and low gas-phase oxygen abundances. All DWALIN galaxies have archival data from the MUSE observations that has a field of view (FoV) of $1'\times1'$, spatial sampling of $0.2''\times0.2''$, wavelength coverage from 4650 to 9300~\AA, and spectral resolution of 1750 (at 4650~\AA) to 3750 (at 9300~\AA). From these 40 galaxies, we first exclude galaxies with distances $>5$ Mpc, to ensure that the full width at half maximum of the spatial point spread function ($\approx 0.9''$ due to seeing) corresponds to a physical length $<20$ pc so that we can resolve characteristic element injection widths. Second, we exclude galaxies with apparent axis ratio $b/a < 0.4$, to minimize corrections for projection effects. Third, we exclude one galaxy (NGC 1068) for which the majority of spaxels are impacted by ionization from an active galactic nucleus. These selections leave NGC 5253 as the only suitable target. NGC 5253 is a metal-poor \citep[$Z \approx 0.2Z_{\odot}$;][]{Monreal-Ibero12a} star-forming dwarf galaxy at a distance of $3.65$ Mpc \citep{Tully16a}, and has an inclination angle of 64$^{\circ}$ \citep[$b/a = 0.43$;][]{Lauberts89}. 

The MUSE data reduction was carried out with the MUSE pipeline \citep{Weilbacher20} v2.8.1 using the \textsc{esorex} pipeline\footnote{\href{https://www.eso.org/sci/software/cpl/esorex.html}{https://www.eso.org/sci/software/cpl/esorex.html}} Common Pipeline Library reduction recipes. The resulting data cubes are then processed with the IDL emission line extraction pipeline \textsc{lzifu}\footnote{\href{https://github.com/hoiting/LZIFU}{https://github.com/hoiting/LZIFU}} to estimate the ionized gas emission line fluxes and uncertainties. We adopt one-component emission line fitting and no spatial smoothing to avoid introducing artificial correlations. Finally, we de-redden the data by comparing the observed Balmer decrements with the intrinsic flux ratio of 2.86 for Case B recombination assuming electron density of 100 cm$^{-3}$ and electron temperature of 10,000 K \citep{Osterbrock89}, and an extinction curve of $R_V=4.05$ \citep{Calzetti00}.

\section{Results}
\label{sec:results}

\subsection{Direct method to estimate chemical abundances}
\label{subsec:abun}

To measure electron temperatures $T_e$ from the reddening-corrected flux ratios we use the fitting results \citep{Perez-Montero17} from the emission-line analysis software \textsc{pyneb} \citep{Luridiana15}. We have verified that alternative diagnostics \citep[e.g.][which gives different coefficients]{Izotov06} produce temperature differences smaller than 1,000 K, leading to only marginal differences in the final derived abundances ($<0.02$ dex).

As MUSE's wavelength range does not cover the collisionally excited doublet [O\textsc{ii}]$\lambda\lambda3727,3729$ or the temperature-sensitive auroral line [O\textsc{iii}]$\lambda4363$, we instead use the collisionally excited line [S\textsc{iii}]$\lambda9069$ and the auroral line [S\textsc{iii}]$\lambda6312$ to estimate electron temperatures in the S$^{++}$ (medium $T_e$) zones and then use those to estimate electron temperatures in the O$^{++}$ (high $T_e$) and O$^{+}$ ($\approx$S$^{+}$, low $T_e$) zones, following the calibration of \cite{Perez-Montero17}:
\begin{equation}
R_{\rm S3} = \frac{I_{9069,9532}}{I_{6312}} \approx \frac{3.44 I_{9069}}{I_{6312}},
\end{equation}
\begin{equation}
t([{\rm S}\textsc{iii}]) = \frac{T([{\rm S}\textsc{iii}])}{10^4~{\rm K}} = 0.5147 + 0.0003187R_{\rm S3} + \frac{23.64}{R_{\rm S3}}, 
\end{equation}
\begin{equation}
t([{\rm O}\textsc{iii}]) = \frac{T([{\rm O}\textsc{iii}])}{10^4~{\rm K}} = 1.19t([{\rm S}\textsc{iii}]) - 0.32,
\end{equation}
\begin{equation}
t([{\rm O}\textsc{ii}]) = \frac{T([{\rm O}\textsc{ii}])}{10^4~{\rm K}} = \frac{2}{t([{\rm O}\textsc{iii}])^{-1} + 0.8},
\label{eqn:t_OII}
\end{equation}
where in the expressions above $I_{9069}$, $I_{9532}$, and $I_{6312}$ are the flux intensities of the [S\textsc{iii}]$\lambda9069$, [S\textsc{iii}]$\lambda9532$, and [S\textsc{iii}]$\lambda6312$ lines, respectively, and $t([{\rm S}\textsc{iii}])$, $t([{\rm O}\textsc{iii}])$, and $t([{\rm O}\textsc{ii}])$ are the electron temperatures in the S$^{++}$-, O$^{++}$-, and O$^{+}$-dominated zones in unit of $10^4$ K, respectively. As the [S\textsc{iii}]$\lambda9532$ emission line is not accessible in MUSE wavelength coverage, we adopt a fixed line ratio of $I_{9532} / I_{9069} = 2.44$ \citep{Vilchez96}. For nitrogen, we consider two different possible methods to infer the electron temperature. For our analysis in the main text we use the collisionally excited line [N\textsc{ii}]$\lambda\lambda6548,6584$ and the auroral line [N\textsc{ii}]$\lambda5755$ to estimate electron temperatures in the N$^+$ (low $T_e$) zones following the calibration of \cite{Perez-Montero17}:
\begin{equation}
R_{\rm N2} = \frac{I_{6548,6584}}{I_{5755}},
\end{equation}
\begin{equation}
t([{\rm N}\textsc{ii}]) = \frac{T([{\rm N}\textsc{ii}])}{10^4~{\rm K}} = 0.6153 - 0.0001529R_{\rm N2} + \frac{35.3641}{R_{\rm N2}},
\label{eqn:t_NII}
\end{equation}
where in the expressions above $I_{6548}$, $I_{6584}$, and $I_{5755}$ are the flux intensities of the [N\textsc{ii}]$\lambda6548$, [N\textsc{ii}]$\lambda6584$, and [N\textsc{ii}]$\lambda5755$ lines, respectively, and $t([{\rm N}\textsc{ii}])$ is the electron temperatures in the N$^+$-dominated zones in unit of $10^4$ K.

While this procedure yields an estimate of the electron temperature in the N$^+$-dominated zone that is fully independent of the oxygen and sulfur lines, the faintness of the [N\textsc{ii}]$\lambda5755$ auroral line means that we only detect $\approx 1,500$ pixels at signal-to-noise ratio $>3$, the minimum we accept for the correlation function analysis described below. For this reason, we also consider an alternative approach of approximating the electron temperature in the N$^+$-dominated zone via that in the O$^+$-dominated zone using the scaling relation from the photoionization models in \cite{Perez-Montero09},
\begin{equation}
t([{\rm N}\textsc{ii}]) = \frac{1.85}{t([{\rm O}\textsc{iii}])^{-1} + 0.72}.
\label{eqn:t_NII_i}
\end{equation}
This gives us an electron temperature map for N that coves as much of the galaxy as for O and S, at the price of assuming a relationship between the temperatures in the O$^{++}$ and N$^+$ zones. In \aref{app:alt} we repeat all of our analysis from the main text using this alternative nitrogen map and show that the qualitative results are the unchanged.

\begin{figure*}
\centering
\includegraphics[width=\linewidth]{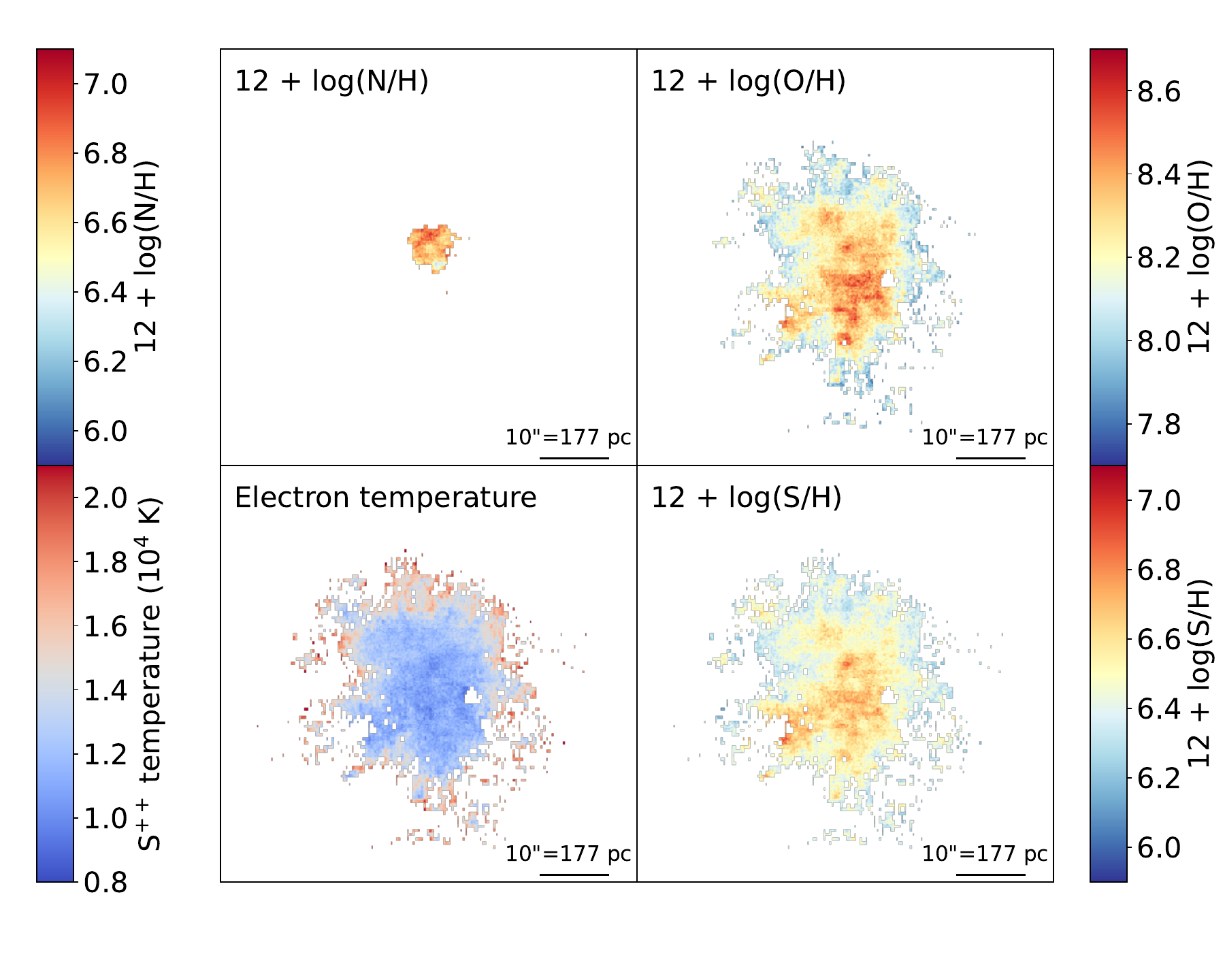}
\caption{[S\textsc{iii}] electron temperature map (lower left) and nitrogen (upper left), oxygen (upper right), and sulfur (lower right) abundance maps for NGC 5253. The fields-of-view are centered on the galactic center and are the same in each panel. The coverage of the nitrogen abundance map is smaller than those of oxygen and sulfur maps because of the faintness of the auroral line [N\textsc{ii}]$\lambda5755$ and the resulting low coverage of electron temperatures in the N$^+$ zones.}
\label{fig:maps}
\end{figure*}

Given the electron temperatures we derive the chemical abundances of O$^+$ and O$^{++}$ from the flux intensities $I_{7320,7330}$ and $I_{4959,5007}$ of the collisionally excited doublets [O\textsc{ii}]$\lambda\lambda7320,7330$\footnote{One may estimate t([{\rm O}\textsc{ii}]) using the line ratio of $I_{7320,7330}/I_{3727,3729}$ \citep[e.g.][]{Osterbrock89, Liang07}. Due to inaccessibility of [O\textsc{ii}]$\lambda\lambda3727,3729$ in MUSE, we infer t([{\rm O}\textsc{ii}]) from t([{\rm O}\textsc{iii}]) as in \autoref{eqn:t_OII}.} and [O\textsc{iii}]$\lambda\lambda4959,5007$, respectively as \cite{Perez-Montero17},
\begin{equation}
\begin{aligned}
12 + \log\left(\frac{\mathrm{O}^+}{\mathrm{H}^+}\right) = & \log\left(\frac{I_{7320,7330}}{I_{\rm H \beta}}\right) + 7.210 + \frac{2.5110}{t([{\rm O}\textsc{ii}])} - 
\\
& 0.422\log[t([{\rm O}\textsc{ii}])] + 0.000398 n_e,
\label{eqn:Z_OII}
\end{aligned}
\end{equation}
\begin{equation}
\begin{aligned}
12 + \log\left(\frac{\mathrm{O}^{++}}{\mathrm{H}^+}\right) = & \log\left(\frac{I_{4959,5007}}{I_{\rm H \beta}}\right) + 6.187 + \frac{1.2491}{t([{\rm O}\textsc{iii}])} -
\\
& 0.582 \log[t([{\rm O}\textsc{iii}])],
\label{eqn:Z_OIII}
\end{aligned}
\end{equation}
where $I_\mathrm{H\beta}$ is the flux intensity of the H$\beta$ line and $n_e$ is the electron density; we assume a value $n_e=100$ cm$^{-3}$ but this choice has minimal effects given the extremely small coefficient. We assume that O$^+$ and O$^{++}$ are the only two ionization states that are significantly populated, and thus the overall oxygen abundance is $12 + \log(\mathrm{O}/\mathrm{H}) = 12 + \log(\mathrm{O}^+/\mathrm{H}^+ + \mathrm{O}^{++}/\mathrm{H}^+)$.
Though N$^+$ is not the only nitrogen ionization state in the nebula, we note that the contribution from N$^{++}$ to the total nitrogen abundance is negligible based on the work in the Milky Way \citep{Pineda24}, and N$^{++}$ emissions are inaccessible in the optical bands. We derive the overall nitrogen abundance $12 + \log(\mathrm{N}/\mathrm{H}) = 12 + \log(\mathrm{N}^+/\mathrm{H}^+)$ from the flux intensities $I_{6548,6584}$ of the [N\textsc{ii}]$\lambda\lambda 6548,6584$ doublet as
\begin{equation}
\begin{aligned}
12 + \log\left(\frac{\mathrm{N}^+}{\mathrm{H}^+}\right) = & \log\left(\frac{I_{6548,6584}}{I_{\rm H \beta}}\right) + 6.291 + \frac{0.9022}{t([{\rm N}\textsc{ii}])} - \\
& 0.5511 \log[t([{\rm N}\textsc{ii}])].
\label{eqn:Z_NII}
\end{aligned}
\end{equation}
Finally, we obtain the chemical abundances of S$^+$ and S$^{++}$ from the flux intensities $I_{6717,6731}$ and $I_{6312}$ of the collisionally excited doublet [S\textsc{ii}]$\lambda\lambda6717,6731$ and auroral line [S\textsc{iii}]$\lambda6312$, respectively as \cite{Perez-Montero17},
\begin{equation}
\begin{aligned}
12 + \log\left(\frac{\mathrm{S}^+}{\mathrm{H}^+}\right) = & \log\left(\frac{I_{6717,6731}}{I_{\rm H \beta}}\right) + 5.463 + \frac{0.941}{t([{\rm O}\textsc{ii}])} - \\
& 0.370\log[t([{\rm O}\textsc{ii}])],
\label{eqn:Z_SII}
\end{aligned}
\end{equation}
\begin{equation}
\begin{aligned}
12 + \log\left(\frac{\mathrm{S}^{++}}{\mathrm{H}^+}\right) = & \log\left(\frac{I_{6312}}{I_{\rm H \beta}}\right) + 6.695 + \frac{1.6640}{t([{\rm S}\textsc{iii}])} - 
\\
& 0.513 \log[t([{\rm S}\textsc{iii}])].
\label{eqn:Z_SIII}
\end{aligned}
\end{equation}
We adopt the approximation of $t([{\rm S}\textsc{ii}]) \approx t([{\rm O}\textsc{ii}])$ \citep{Perez-Montero17}. The overall sulfur abundance is then $12 + \log(\mathrm{S}/\mathrm{H}) = 12 + \log(\mathrm{S}^+/\mathrm{H}^+ + \mathrm{S}^{++}/\mathrm{H}^+)$, similarly assuming that S$^+$ and S$^{++}$ are the two dominant ionization states \citep{Amayo21}. The [S\textsc{iii}] electron temperature map and the abundance maps for each element are shown in \autoref{fig:maps}.

\subsection{Two-point autocorrelation function and model fit}
\label{subsec:tpcf}

After having abundance maps we compute their two-point autocorrelation and cross-correlation functions following \cite{Li21, Li23}. We start from the chemical abundance map $Z_{\mathrm{X}} = 12 + \log$(X/H) for elements X $=$ O, N, or S, masking out any pixels for which any of the lines required to derive the abundance fall below a minimum signal-to-noise of 3; \cite{Li21} show that applying a signal-to-noise cut of this form is required to suppress noise in the correlation function.


We subtract the radial abundance gradients for each element to produce residual maps. To do so we compute the average abundance, $\overline{Z}_{\mathrm{X}}$, in annular bins centered on the galactic center and with a width equal to the sampling scale ($0.2''$) of the MUSE instrument, which corresponds to 3.5 pc at the distance of NGC 5253. We then subtract the radial averages from the individual pixel abundances $Z_{\mathrm{X},i}$ to obtain residual abundances $Z'_{\mathrm{X},i} = Z_{\mathrm{X},i} - \overline{Z}_{\mathrm{X}}$. The residual maps have zero mean and no first-order radial gradients by construction.

We compute the two-point autocorrelation functions of the abundance residual maps for each element, averaged over a range of separations from $r_n$ to $r_{n+1}$,
\begin{eqnarray}
\xi_{\mathrm{XX}, n} & = & \frac{\sigma_{Z_\mathrm{X}'}^{-2}}{N_n} \sum_{r_n < r_{ij} \leq r_{n+1}} Z'_{\mathrm{X},i} Z'_{\mathrm{X},j},
\\
\sigma_{Z'_\mathrm{X}}^2 & = & \frac{1}{N_p} \sum_{i=1}^{N_p} {Z'_{\mathrm{X},i}}^2,
\end{eqnarray}
where the sum runs over the $N_n$ pixel pairs $(i,j)$ for which $r_n < |\mathbf{r}_{i} - \mathbf{r}_j| \leq r_{n+1}$, and $\sigma_{Z_\mathrm{X}'}^2$ is the total variance of the $N_p$ pixels in the residual map for element X (typically $N_p\gtrsim25,000$ for oxygen and sulfur abundances, and $N_p\approx1,500$ for nitrogen abundances). The pixel number and area coverage for each element abundance map reach the minimum requirements proposed in \cite{Li21}. As with the gradient subtraction step, we use bins 3.5 pc wide (the native sampling rate of the data cube) to a maximum separation of 600 pc.

We account for observational uncertainty in the two-point autocorrelation function by performing 50 bootstrap trials. In each trial, we generate a line flux map by sampling each pixel $i$ from the normal distribution with mean $f_i$ and standard deviation $\sigma_{f, i}$ to generate a new map. For the bootstrapped map we use the direct method described above to compute the abundance, then compute the two-point autocorrelation. The mean and standard deviation across the trials provide the final estimate and uncertainty. We note that the uncertainties estimated in this way only take measurement errors into account, and are sub-dominant compared to the much larger systematic uncertainties arising from the assumptions embedded in the line diagnostics used to convert fluxes to metallicity measurements; these are unfortunately very difficult to quantify.

We perform a parametric fit to the derived two-point autocorrelation function using the functional form proposed by \cite{KT18}, along with the modifications to account for observational uncertainties described in \cite{Li21}. The model treats the metal field as the result of a stochastically-forced diffusion process, and takes the functional form
\begin{equation}
\begin{aligned}
\xi_{\rm model}(r) = & \int_0^\infty e^{-\sigma^2 a^2} \left(1 - e^{-2 l^2 a^2}\right) \frac{J_0(ar)}{a} \, da \times \\
& \frac{2}{\ln\left(1 + 2 l^2 / \sigma^2 \right)} \times \frac{1}{f}, (r > 0)
\\
\sigma^2 = & \frac{1}{2}\sigma_{\rm beam}^2 + w_{\rm inj}^2,
\label{eqn:model}
\end{aligned}
\end{equation}
where $\sigma_{\rm beam}$ is fixed at the observational beam in physical distance ($\sim6.8$ pc in our case, corresponding $1/2.354$ of the typical seeing $0.9''$ at the distance of NGC 5253), $w_{\rm inj}$ is the injection width parameter describing the initial radius of the region into which metals are injected, $l$ is the correlation length set by turbulent mixing of metals in the ISM, $f$ is the factor by which observational uncertainties in the derived chemical abundances increase the variance in the abundance residuals compared to the true variance, and $J_0(x)$ is the Bessel function of the first kind of order zero.

\begin{figure*}
\centering
\includegraphics[width=\linewidth]{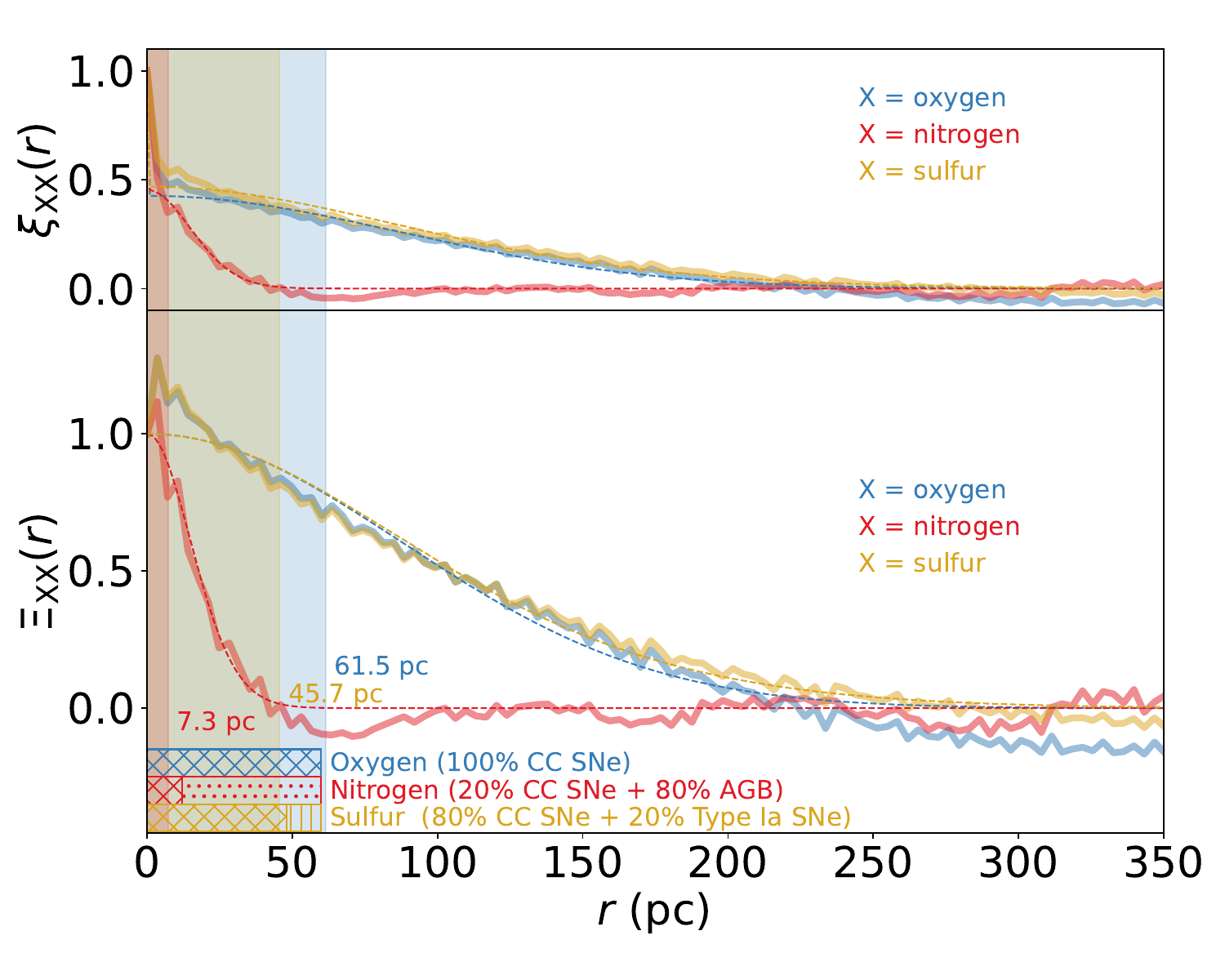}
\caption{Two-point autocorrelation functions (thick solid lines) for oxygen (blue), nitrogen (red), and sulfur (yellow) and our best-fit parametric models (thin dashed curves). In both panels correlation $=0$ means uncorrelated and $=1$ means perfectly correlated. The upper panel shows the measured autocorrelation, while the lower panel shows the true autocorrelation corrected by our median estimate for the factor by which the autocorrelation is reduced by measurement uncertainties, to focus on the shape of the curve, which is what controls the fit for the injection width parameter. The stripes on the left represent the values of injection widths with each element shown in the same colors as their corresponding elements. In the lower panel we show the enrichment source contribution fractions using three horizontal bars \citep{Rybizki17} in the same colors as their corresponding elements. The fitted injection width parameters for oxygen (the blue stripe) and sulfur (the yellow stripe) agree well with the characteristic maximum radii of supernova blast waves in the ISM from both numerical simulations \citep[59 pc;][]{Kolborg22} and theoretical models \citep[67 pc;][]{Draine11}. The fitted injection width parameters for nitrogen (the red stripe) is much smaller than those for oxygen and sulfur, suggesting AGB stars as dominant origin sites.}
\label{fig:auto_corr}
\end{figure*}

\begin{figure*}
\centering
\includegraphics[width=\linewidth]{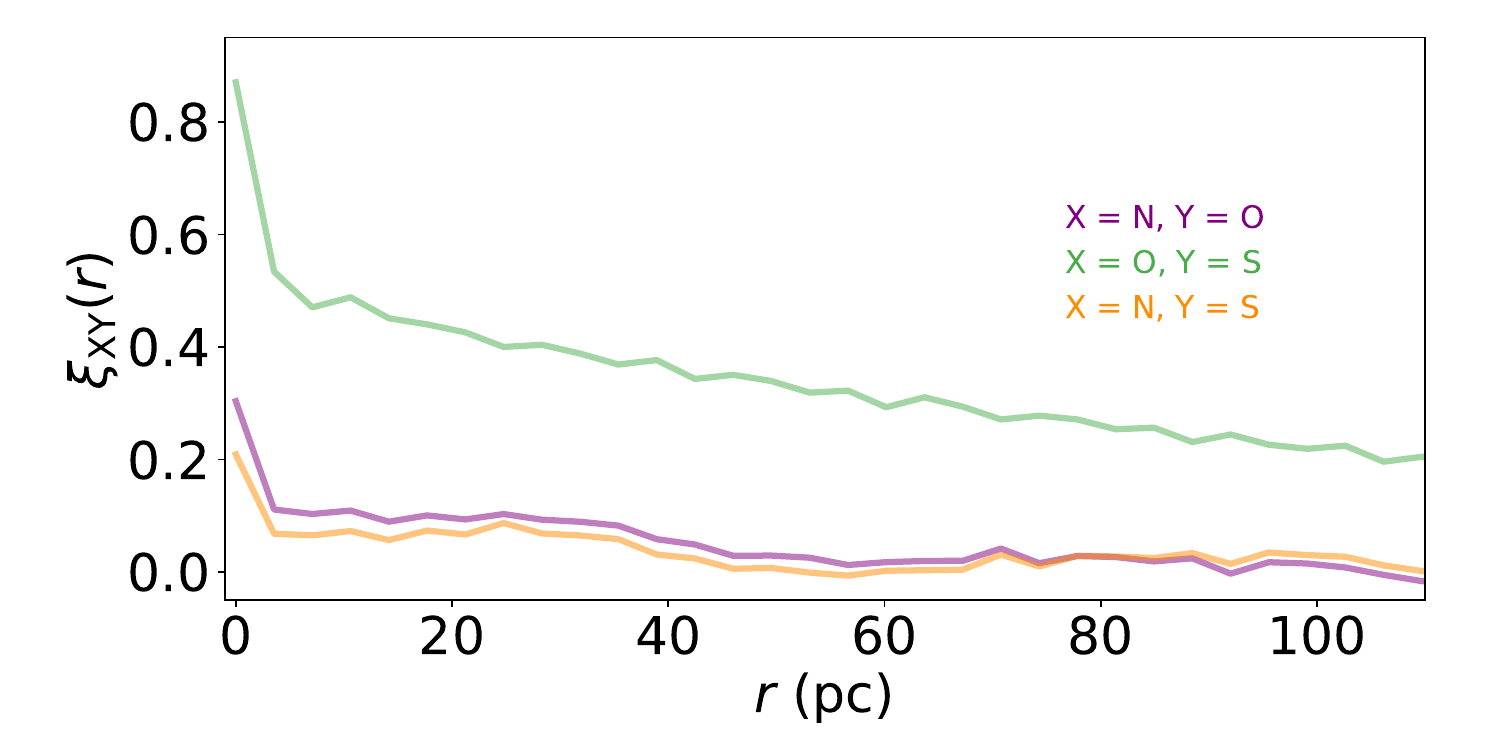}
\caption{Cross-correlation functions for N-O (purple), O-S (green), and N-S (orange), uncorrected for observational uncertainties. Correlation increases from lower (uncorrelated) to upper (nearly perfectly correlated). At scales smaller than 60 pc, the O-S cross-correlation functions are most correlated, followed by those of N-O and N-S, thanks to the same origin site (core-collapse supernovae) that oxygen and sulfur share.}
\label{fig:cross_corr}
\end{figure*}

To derive values for our model parameters -- $w_\mathrm{inj}$, $l$, and $f$ -- for each element, we carry out MCMC fits of this model to the measured two-point autocorrelation functions (including their uncertainties). These fits use 100 walkers and 5,000 total steps, with the first 4,000 used for burn-in and the posterior PDF derived from the final 1,000 steps. See \cite{Li21, Li23} for more details on the fitting method. We show the two-point autocorrelation functions and parametric fits of $w_{\rm inj}$ for the O, N, and S abundance maps in \autoref{fig:auto_corr}. The full posterior PDFs for all fitted quantities are shown in \aref{app:pdf}.

A key output of the model fits to the autocorrelation functions are the injection widths of the oxygen, nitrogen, and sulfur abundance maps, defined as the characteristic sizes of the regions into which injection events deposit these elements before they begin to be distributed by ISM turbulence. Our fits return the median and 90\% confidence intervals of injection widths $w_\mathrm{inj} = 61.5\pm0.3$, $7.3\pm0.2$, and $45.7\pm0.7$ pc for the O, N, and S abundance maps, respectively. The similar values of $w_\mathrm{inj}$ for oxygen and sulfur are in excellent agreement with the characteristic maximum radii of SN blast waves in the ISM from both numerical simulations \citep[e.g. 59 pc in low-star-formation-rate regions;][]{Kolborg22} and theoretical models \citep[e.g. 67 pc given SN explosion energy of $10^{51}$ erg, ISM density of 1 hydrogen atom per cm$^3$, and ISM velocity dispersion of 10 km s$^{-1}$;][]{Draine11}. This radius is a factor of $\sim 4$ larger than the typical observed radii of SN remnants \citep{Gao24} and surrounding gas of Wolf-Rayet stars \citep{Lopez-Sanchez11, Sarbadhicary23}, possibly because such studies are biased to select SN remnants when they are smaller and therefore brighter. Our measurement for injection widths suggests that two-point autocorrelations of oxygen and sulfur abundance distributions in galaxies directly encode SN remnant sizes, making it possible to infer these sizes even in galaxies that lack recent SNe.

By contrast, the injection width for nitrogen is a factor of $\approx 8$ smaller, a difference that is detected at very high statistical significance ($\gg10\sigma$). This is consistent with this element being injected by a mixture of two processes: a dominant ($\approx 80\%$, as computed with \textsc{ChemPy} \citep{Rybizki17}) contribution from AGB stars, with very small injection widths due to the low energy with which material is ejected, along with a subdominant ($\approx 20\%$) contribution from CC SNe. Our result demonstrates that differences in astrophysical origin leave detectable imprints on the spatial distributions of elemental abundances in the ISM\footnote{Some evidence \citep[e.g.][]{Vangioni18, Vincenzo18} suggests that the increasing production of N at high O abundance is simply the result of metal-dependent SN yields in addition to``failed SNe'', but out of the scope of this work.}.

\begin{figure*}
\centering
\includegraphics[width=\linewidth]{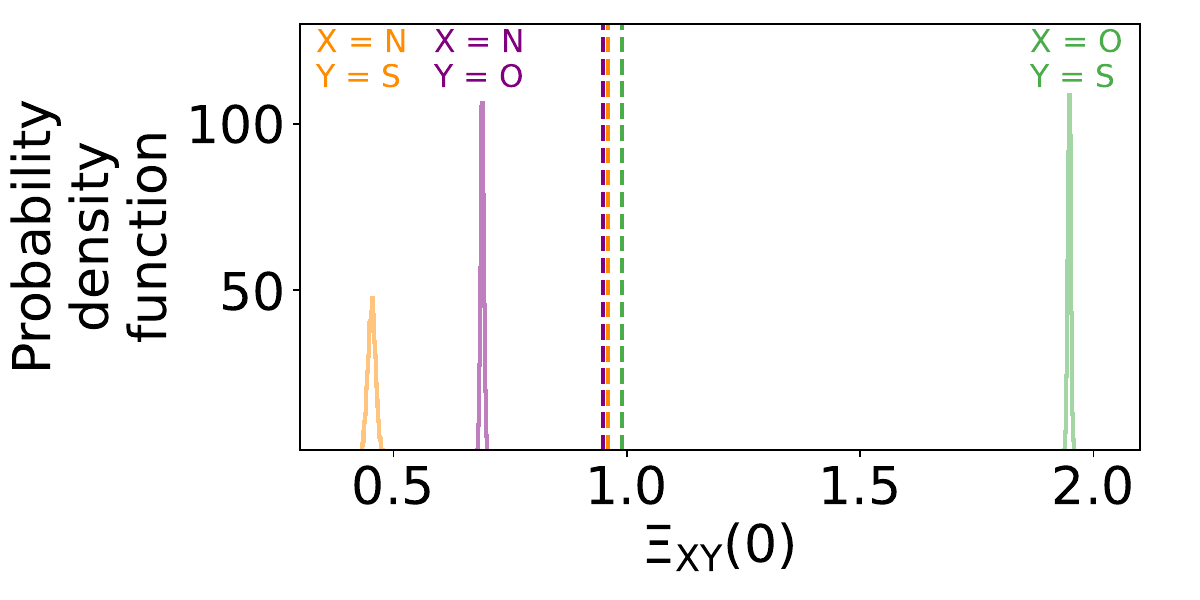}
\caption{Posterior PDFs of true zero-lag cross-correlations $\Xi_{\rm XY}(0)$ for each pair of elements, corrected for observational uncertainties. Correction increases from left to right. The vertical dashed lines indicate the zero-lag cross-correlations predicted by simulations \citep{Zhang25}. Our estimated zero-lag cross-correlations suggest that all the elements are correlated ($>0.4$) in the spatial distributions of their abundances, among which oxygen and sulfur are the most correlated pair as they have the same origin site.}
\label{fig:pdf_cross}
\end{figure*}

\subsection{Cross-correlation function}
\label{subsec:ccf}

The first half of the procedure to derive element-element cross-correlation functions is identical to that for the autocorrelation functions: we produce abundance residual maps, $Z'_{\mathrm{X}, i}$ and $Z'_{\mathrm{Y}, i}$ for elements X and Y exactly as described above. We then compute the bin-averaged cross-correlation function as
\begin{eqnarray}
\xi_{\mathrm{XY}, n} & = & \frac{(\sigma_{Z'_{\rm X}}\sigma_{Z'_{\rm Y}})^{-1}}{N_n} \sum_{r_n < r_{ij} \leq r_{n+1}} Z'_{\mathrm{X}, i} Z'_{\mathrm{Y}, j},
\end{eqnarray}
using the same radial bins as for the autocorrelation analysis, and considering only pixels where both elements are detected. We show the cross-correlation functions $\xi_{\rm NO}(r)$, $\xi_{\rm OS}(r)$, and $\xi_{\rm NS}(r)$ in \autoref{fig:cross_corr}. We derive uncertainties on the cross-correlation functions using the same bootstrapping process discussed above for autocorrelation functions.

We are particularly interested in the cross-correlation at zero lag, which we denote $\Xi_{\rm XY}(0)$ for convenience. Estimating this quantity requires some care; a naive estimate would be simply to set this equal to $\xi_{\mathrm{XY}}(0)$, the cross-correlation we measure at zero lag, but the presence of observational uncertainties complicates this approach. We obtain $\xi_{\mathrm{XX}}(0) = 1$ by construction, even in the presence of uncertainties in measuring the element abundance, because the uncertainties in the abundance of a particular element in a particular pixel are perfectly correlated with themselves. The uncertainties of two different elements, however, are not, and thus the presence of measurement uncertainties in the chemical abundance maps means that the cross-correlation in the measured map is lower than it is in reality.

If the uncertainties in the two abundance maps were uncorrelated, and the effect of observational uncertainties on each map is simply to increase the variance (as assumed in our models of the autocorrelation function), then the true and measured zero-lag cross-correlations are related by 
\begin{eqnarray}
\Xi_{\rm XY}(0) = \sqrt{f_{\rm XX} f_{\rm YY}} \xi_{\rm XY}(0),
\end{eqnarray}
where $f_{\rm XX}$ and $f_{\rm YY}$ are the factors by which measurement uncertainties increase the variance of each of the element abundance maps. We therefore derive our final estimates of $\Xi_{\rm XY}(0)$ from the posterior PDFs of $f_{\rm XX}$ and $f_{\rm YY}$ derived from the MCMC fits using bootstrapping: we randomly draw values from the posterior PDFs of $f_{\rm XX}$ and $f_{\rm YY}$, and values for $\xi_{\rm XY}(0)$ from the Gaussian distribution derived for the zero-lag bin of the cross-correlation. For each realization we compute $\Xi_{\rm XY}(0)$; we show the PDF of samples returned by this process in \autoref{fig:pdf_cross}, and derive our median and confidence intervals for the zero-lag cross correlation from these samples.

The recovered O-S zero-lag cross-correlation $\Xi_{\rm OS}(0)$ is significantly larger than $\Xi_{\rm NO}(0)$ and $\Xi_{\rm NS}(0)$, which in turn are close to equal to one another. Numerically, we obtain the median and 90\% confidence intervals of cross-correlations at zero lag $\Xi_{\rm NO}(0)=0.691\pm0.007$, $\Xi_{\rm OS}(0)=1.949\pm0.006$, and $\Xi_{\rm NS}(0)=0.456\pm0.014$. Note that values $>1$ are possible due to the approximations involved in correcting for observational uncertainties (see \aref{app:ccf}). Our finding is supported by a recent prediction from simulations \citep{Zhang25} that N-O and N-S cross-correlations at zero lag should be smaller than O-S ones because that element-element cross-correlations break up into groups of different nucleosynthetic origins: elements that arise from the same astrophysical production site are nearly perfectly correlated, while those from different production sites are significantly less correlated. Our measurement in NGC 5253 provides strong observational confirmation of this prediction. However, the direct comparison between the results of this work and the \citeauthor{Zhang25} simulations is limited, as the simulations did not include older AGB stars as they only run $\sim500$ Myr, and so was missing a potentiallt significant source of nitrogen. Zhang et al. (in prep., priv. comm.) also find that stellar clustering in the simulations is too strong, in the sense that most stars stay together in bound clusters for at least a few hundred Myr, contrary to observations. This may elevate the correlations above what they should be, and is probably down to the technical details of pre-SN stellar feedback.

\section{summary and conclusions}
\label{sec:conclusions}

Our results confirm the feasibility of using elemental abundance correlation functions to reveal the relations between the origin sites of chemical elements and the extragalactic spatial distributions of their abundances -- autocorrelation functions offer insights into the impact radii of SN explosions and the multiple enrichment sources of nitrogen, while cross-correlation functions highlight the underlying connections between the enrichment sources of elements. This work also has important implications for studies on stellar abundances that attempt to use these abundances to reconstruct star clusters -- the so-called ``chemical tagging'' method \citep{Casamiquela21}. Our results suggest that element abundances in the gas phase are highly structured by nucleosynthetic group, with high internal correlations within a group, indicating that stellar chemical abundances are likely to be decomposable into a small number of components that explain most of the variation. This confirms recent suggestions \citep{Ness22a, Ting22a, Griffith24a} that there are fundamental limits on the amount of information that can be recovered from stellar abundances.

\begin{acknowledgments}
ZL and AMS acknowledge the Science and Technology Facilities Council (STFC) consolidated grant ST/X001075/1. MRK acknowledges support from the Australian Research Council through Laureate Fellowship FL220100020.
\end{acknowledgments}

\begin{contribution}

ZL conducted the analysis, made the plots, and led the writing of the preliminary version of the draft. MRK, AFM, and AMS suggested the physical pictures discussed in this paper and edited the manuscript. EW, JTM, FB, and GC were involved in the comments in the manuscript and the interpretation of the results. FB, GC, GV, and JK conducted the data reduction.


\end{contribution}

%



\appendix

\section{Alternative nitrogen abundance estimates}
\label{app:alt}

As mentioned in \autoref{subsec:abun}, in the main text we estimate the electron temperature in the N$^+$ zone t([N\textsc{ii}]) from the auroral line [N\textsc{ii}]$\lambda5755$. This lets us measure t([N\textsc{ii}]) that is fully independent from those in the O$^{++}$ zone t([O\textsc{iii}]) and in the S$^{++}$ zones t([S\textsc{iii}]), but at the cost of producing a nitrogen abundance map that covers significantly less of the galaxy than the O or S abundance maps due to the faintness of [N\textsc{ii}]$\lambda5755$. To investigate whether this smaller coverage area biases our results, here we investigate the alternative approach of measuring the nitrogen abundance instead using t([N\textsc{ii}]) inferred from t([O\textsc{iii}]) given the scaling relation in \autoref{eqn:t_NII_i} \citep{Perez-Montero09, Perez-Montero17}. Note that this step is used only for the purposes of estimating electron temperatures in the N$^+$ zones; the actual nitrogen abundances are still derived from the collisionally excited  lines that are independent of the lines used to measure O or S, rather than from an assumed N/O scaling as assumed in strong-line methods.

\begin{figure*}
\centering
\includegraphics[width=\linewidth]{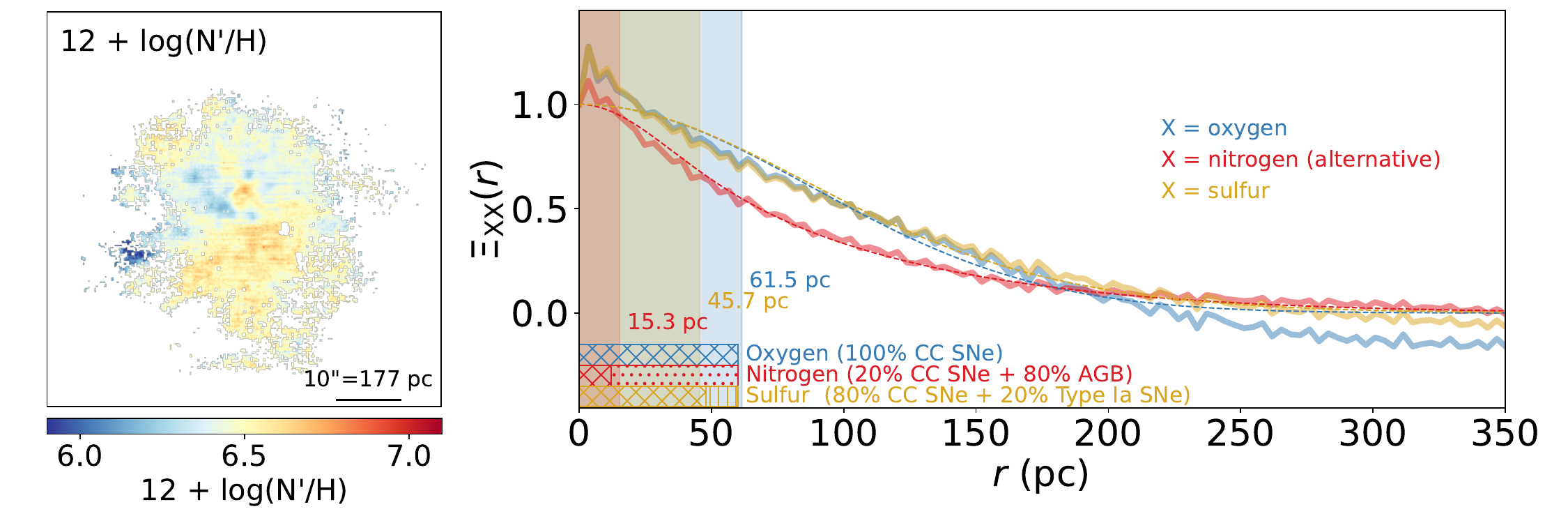}
\caption{Alternative versions of the upper left panel of \autoref{fig:maps} (left) and the lower panel of \autoref{fig:auto_corr} (right), where nitrogen abundance is measured using the inferred electron temperatures in the N$^+$ zone (\autoref{eqn:t_NII_i}). The conclusion remains unchanged that the initial injection for nitrogen ($15.3\pm0.2$ pc) occurs on significantly smaller scales than those for oxygen and sulfur.}
\label{fig:nit_alt_auto}
\end{figure*}

\begin{figure*}
\centering
\includegraphics[width=\linewidth]{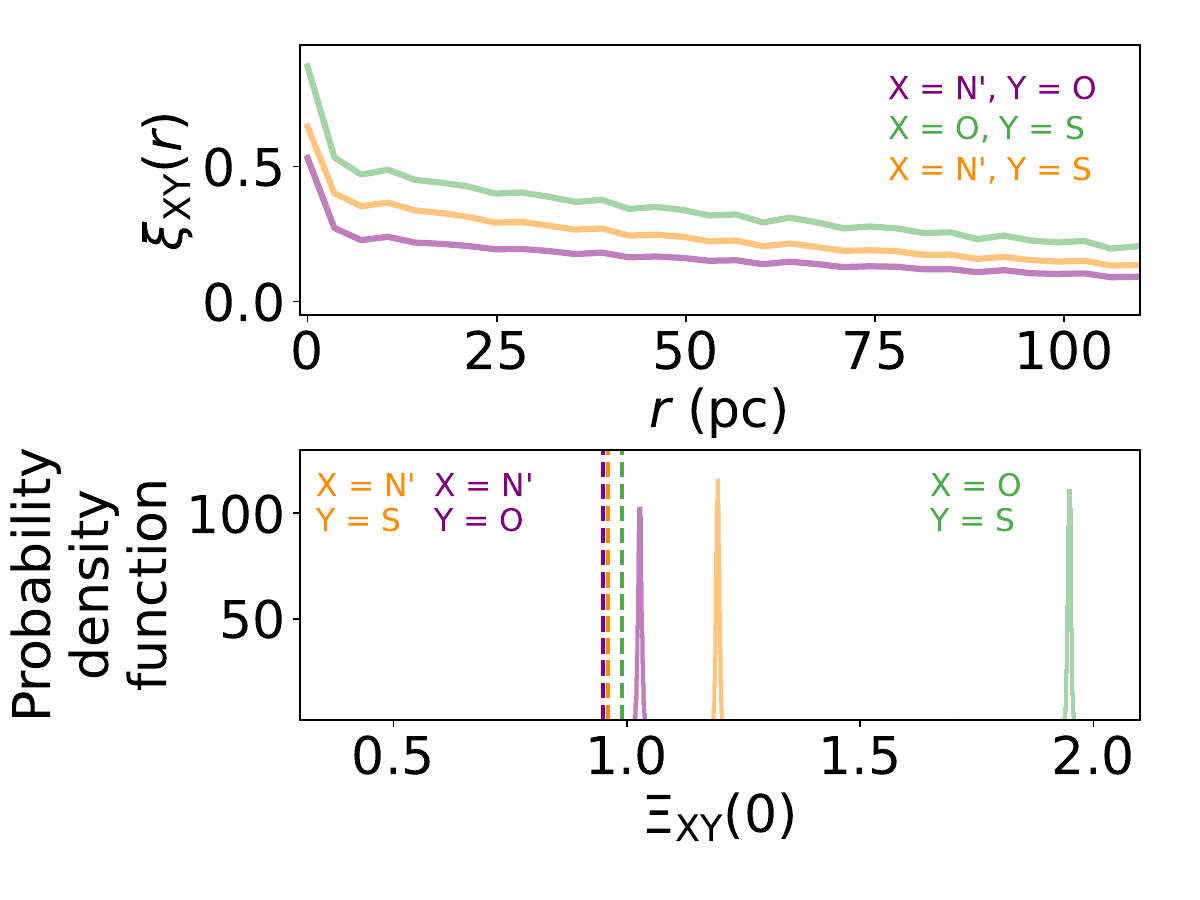}
\caption{Alternative versions of \autoref{fig:cross_corr} (upper) and \autoref{fig:pdf_cross} (lower), where nitrogen abundance is measured using the inferred electron temperatures in the N$^+$ zone (\autoref{eqn:t_NII_i}). The conclusion remains unchanged that oxygen and sulfur are better correlated with each other than with nitrogen [$\Xi_{\rm N'O}(0) = 1.028\pm0.007$ and $\Xi_{\rm N'S}(0) = 1.195\pm0.06$].}
\label{fig:nit_alt_cross}
\end{figure*}

We repeat all the analysis steps described in \autoref{sec:results} using this alternative nitrogen abundance map, and in \autoref{fig:nit_alt_auto} and \autoref{fig:nit_alt_cross} we reproduce \autoref{fig:maps} - \autoref{fig:pdf_cross} of the main text for this alternative estimate of N abundance, which we denote N$'$. For this map we obtain the median and 90\% confidence intervals for the autocorrelation fit parameters [$w_{\rm inj}$, $l$, $f$] $=$ [$15.30\pm0.08$ pc, $84.16\pm0.11$ pc, $1.594\pm0.002$], and true cross-correlations at zero lag $\Xi_{\rm N'O}=1.028\pm0.007$ and $\Xi_{\rm N'S}=1.195\pm0.006$.

Our analysis shows that the results remain quantitatively stable and conclusions do not change for this alternative N map. Comparing the right panel of \autoref{fig:nit_alt_auto} to \autoref{fig:auto_corr}, we see that the qualitative difference in shape between the N autocorrelation function and the O and S ones remains the same, and retain the key property that $w_\mathrm{inj}$ is much smaller for N than for O or S. Comparing the cross-correlation functions of $\xi_{\rm N'Y}(r)$ to $\xi_{\rm NY}(r)$ for Y $=$ O or S (the upper panel of \autoref{fig:nit_alt_cross} versus \autoref{fig:cross_corr}), and the values of $\Xi_{\rm N'Y}(0)$ to $\Xi_{\rm NY}(0)$ (the lower panel of \autoref{fig:nit_alt_cross} versus \autoref{fig:pdf_cross}) we again see that the cross-correlation of N with O or S is much smaller than the cross-correlation of O with S. Thus our analysis shows that qualitative results do not depend on the N coverage map.

\section{Full posterior probability distribution functions of the fitted parameters}
\label{app:pdf}

\begin{figure*}
\centering
\includegraphics[width=\linewidth]{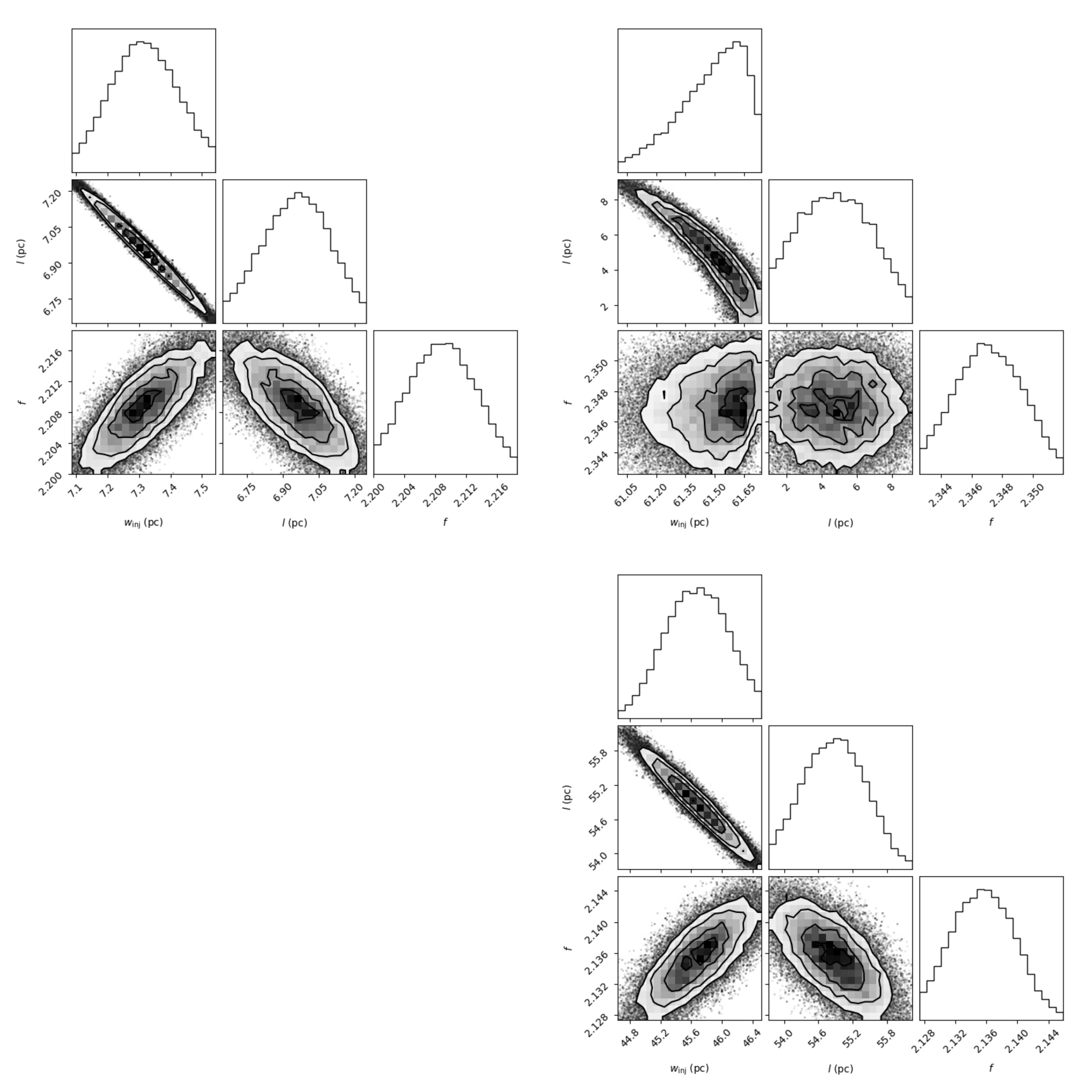}
\caption{The distributions of our three fit parameters (injection width $w_{\rm inj}$, correlation length $l$, and the factor by which the autocorrelation is reduced by measurement uncertainties $f$), describing the autocorrelation functions of nitrogen (upper left), oxygen (upper right), and sulfur (lower right) measured in NGC 5253 as derived from our MCMC. In each panel, the heat map colors black, dark gray, light gray, and white (i.e. the four contours from inside to outside) show probability densities corresponding to 1, 2, 3, and 4$\sigma$, respectively. Partially transparent points show individual samples in regions of lower probability. We list the median and 90\% confidence intervals of the fitted parameters [$w_\mathrm{inj}$, $l$, $f$] $=$ [$61.5\pm0.3$ pc, $4.8\pm3.6$ pc, $2.347\pm0.004$] for the O abundance map, $=$ [$7.3\pm0.2$ pc, $7.0\pm0.2$ pc, $2.209\pm0.008$] for the N abundance map, and $=$ [$45.7\pm0.7$ pc, $54.9\pm1.0$ pc, $2.135\pm0.007$] for the S abundance map.}
\label{fig:corner}
\end{figure*}

Our fits return the median and 90\% confidence intervals of injection widths $w_\mathrm{inj} = 61.5\pm0.3$, $7.3\pm0.2$, and $45.7\pm0.7$ pc for the O, N, and S abundance maps, respectively (\autoref{fig:corner}).

\section{Super-unity zero-lag cross correlation}
\label{app:ccf}

Note that, in \autoref{subsec:ccf} after correcting for observational uncertainties, our O-S zero-lag cross correlation is larger than unity, which is mathematically forbidden. This is a result of our correction somewhat overshooting because we assume that the uncertainties on different elements are uncorrelated, which is not precisely correct as the use of a common electron temperature scale for O and S induces correlations between them. Thus any uncertainties we make in deriving the electron temperature will affect the derived abundances in correlated ways. This unaccounted-for correlation is almost certainly the reason our correction for measurement uncertainties overestimates and returns cross-correlations $>1$. This hypothesis is supported by our analysis using the alternative N map in \aref{app:alt}. There we show that, if we calculate N$^+$ abundances using an electron temperature scaled from that using the same auroral line as for oxygen, the N-O and N-S zero-lag cross-correlations, while still remaining much smaller than for O-S, rise significantly and become greater than unity.


\bibliography{ref}{}
\bibliographystyle{aasjournalv7}

\end{document}